\documentstyle[prl,aps,twocolumn]{revtex}

\newcommand{\eq}{\begin{equation}}
\newcommand{\en}{\end{equation}}
\newcommand{\eqa}{\begin{eqnarray}}
\newcommand{\ena}{\end{eqnarray}}

\newcommand{\la}{\langle}
\newcommand{\ra}{\rangle}
\def\tr{\mbox{tr\,}}
\def\al{\alpha}
\def\be{\beta}
\def\ot{\otimes}

\def\tit#1#2#3#4#5{{#1} {\bf #2}, #3 (#4).}
\def\ti#1#2#3#4#5{{#1} {\bf #2}, #3 (#4)}

\def\prl{Phys. Rev. Lett.}

\def\prb{Phys. Rev. B}

\def\zpb{Z. Phys. B}
\def\ssc{Solid State Comm.}

\begin{document}
\draft

\twocolumn[\hsize\textwidth\columnwidth\hsize\csname @twocolumnfalse\endcsname

\title{Ground state properties of a fully frustrated quantum spin system}

\author{Elliott H. Lieb and Peter Schupp}

\address{Department of Physics, Princeton University,
Princeton, New Jersey 08544}


\maketitle

\begin{abstract}
We find that ground states of the quantum Heisenberg antiferromagnet
on the geometrically frustrated pyrochlore checkerboard lattice are singlets 
and can be expressed in terms
of positive matrices.
The magnetization at zero external field 
vanishes for each 
frustrated tetrahedral unit separately 
and there is an upper bound of 1/8 in natural units on the susceptibility 
both for the ground state and at finite temperature.
These results are the first exact ones in this field and 
generalize to some other lattices; the approach is also of interest 
for other spin systems.
\end{abstract}


]

Geometrically frustrated spin systems are known to have many interesting
properties that are quite unlike those of conventional magnetic systems
or spin glasses~\cite{HTD}. 
Most results are for classical systems.
The first frustrated system, for which the richness
of classical ground states was noted, was the triangular lattice \cite{WH}.
Subsequently, the pyrochlore lattice,
which consists of tetrahedra that share sites, was identified as a
lattice on which the frustration effects are especially strong~\cite{PWA}.
Unusual low-energy properties -- in particular the absence of ordering at
any temperature, was predicted both for discrete \cite{PWA} 
and continuous \cite{JV} classical
spin sytems. The ground state and low energy properties of the classical 
pyrochlore antiferromagnet
-- whose quantum version is focus of this letter --
has been extensivly studied in \cite{MC}.
Both the interest and difficulty in studying frustrated spin systems stem
from the large ground-state degeneracy, which precludes most perturbative
approaches. 

As is the case for most other strong interacting systems in
more than one-dimension, very little is known exactly about the ground states of
frustrated quantum spin systems. Most of the present knowledge has been
obtained by numerics or clever approximations.
Quantum fluctuations
have been studied in the limts of large-$S$ \cite{SL},
where a tendency towards lifting the ground-state degeneracy in favor of
an ordered state (``quantum order by disorder") was detected. In the
opposite limit -- $S=1/2$, where quantum fluctuations are much stronger --
the pyrochlore afm has been identified as a candidate for a quantum disordered
magnet (``quantum spin liquid") \cite{CL}, and it has also been discussed in
terms of an RVB approach \cite{RVB}. However, there are no exact results 
against which to test
the reliability of the results in this limit. In contrast to this, for
conventional -- bipartite -- 
antiferromagnetic spin systems it is well known, for
example, that the energy levels are ordered in a natural
way according to spin, starting from spin zero \cite{LM}. 
Geometrically frustrated systems are not
bipartite and thus this otherwise quite general
theorem does not apply.

In this letter, we find some first exact results for the fully
frustrated quantum antiferromagnet on a pyrochlore checkerboard, 
with the help of the reflection symmetry of this two dimensional lattice.
We shall establish rigorously that 
there is always a ground state with total spin zero. Furthermore, 
in the periodic case
all ground states (if there is more than one) 
have total spin zero, the spin-expectation vanishes for
each frustrated unit (this is the quantum analog of the 
``{\em ice rule}'' for the Ising
system), and there are concrete upper bounds on the susceptibility.

Geometric frustration occurs typically for spin systems with
interactions that favor anti-alignment and involve 
fully connected units of three or more spins, that can obviously not all be 
mutually anti-aligned.
The kagome lattice is an example of frustrated spin systems
with site-sharing triangular units, the pyrochlore lattice
and its two-dimensional version, the pyrochlore checkerboard are examples with
site-sharing tetrahedra. 
We shall focus on
the latter; it is a 
two dimensional array of site-sharing tetrahedra, whose projection 
onto a plane is a square lattice with two
extra diagonal bonds on every other square. (The regular pyrochlore lattice is
a three-dimensional array of site-sharing tetrahedra.)
The tetrahedra -- or squares with extra diagonal bonds -- are 
the frustrated units and will henceforth be called {\em boxes}.

\begin{figure}
\unitlength 1.00mm
\linethickness{0.4pt}
\begin{picture}(77.00,33.00)
\put(6.00,2.00){\line(0,1){30.00}}
\put(12.00,32.00){\line(0,-1){30.00}}
\put(18.00,2.00){\line(0,1){30.00}}
\put(24.00,32.00){\line(0,-1){30.00}}
\put(30.00,2.00){\line(0,1){30.00}}
\put(36.00,32.00){\line(0,-1){30.00}}
\put(3.00,5.00){\line(1,0){36.00}}
\put(39.00,11.00){\line(-1,0){36.00}}
\put(3.00,17.00){\line(1,0){36.00}}
\put(39.00,23.00){\line(-1,0){36.00}}
\put(3.00,29.00){\line(1,0){36.00}}
\put(6.00,29.00){\line(1,-1){24.00}}
\put(6.00,17.00){\line(1,-1){12.00}}
\put(18.00,29.00){\line(1,-1){18.00}}
\put(30.00,29.00){\line(1,-1){6.00}}
\put(6.00,23.00){\line(1,1){6.00}}
\put(6.00,11.00){\line(1,1){18.00}}
\put(12.00,5.00){\line(1,1){24.00}}
\put(24.00,5.00){\line(1,1){12.00}}
\put(21.00,1.00){\line(0,1){2.00}}
\put(21.00,4.00){\line(0,1){2.00}}
\put(21.00,7.00){\line(0,1){2.00}}
\put(21.00,10.00){\line(0,1){2.00}}
\put(21.00,13.00){\line(0,1){2.00}}
\put(21.00,16.00){\line(0,1){2.00}}
\put(21.00,19.00){\line(0,1){2.00}}
\put(21.00,22.00){\line(0,1){2.00}}
\put(21.00,25.00){\line(0,1){2.00}}
\put(21.00,28.00){\line(0,1){2.00}}
\put(21.00,31.00){\line(0,1){2.00}}
\put(75.00,24.00){\line(-1,0){14.00}}
\put(61.00,10.00){\line(1,0){14.00}}
\put(75.00,10.00){\line(-1,1){14.00}}
\put(61.00,10.00){\line(1,1){14.00}}
\put(68.00,15.00){\line(0,1){4.00}}
\put(68.00,22.00){\line(0,1){4.00}}
\put(68.00,29.00){\line(0,1){4.00}}
\put(68.00,5.00){\line(0,-1){4.00}}
\put(61.00,10.00){\line(0,1){2.00}}
\put(61.00,14.00){\line(0,1){2.00}}
\put(61.00,18.00){\line(0,1){2.00}}
\put(61.00,22.00){\line(0,1){2.00}}
\put(75.00,10.00){\line(0,1){2.00}}
\put(75.00,14.00){\line(0,1){2.00}}
\put(75.00,18.00){\line(0,1){2.00}}
\put(75.00,22.00){\line(0,1){2.00}}
\put(59.00,24.00){\makebox(0,0)[rb]{$\bf s_1$}}
\put(59.00,10.00){\makebox(0,0)[rt]{$\bf s_2$}}
\put(77.00,10.00){\makebox(0,0)[lt]{$\bf s_3$}}
\put(77.00,24.00){\makebox(0,0)[lb]{$\bf s_4$}}
\put(68.00,12.00){\line(0,-1){4.00}}
\end{picture}
\caption{(a) pyrochlore checkerboard, reflection symmetric 
about dashed line
(b) frustrated unit with crossing bonds}
\label{fig:che1}
\end{figure}
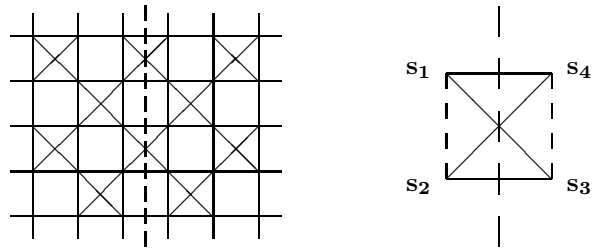 

The hamiltonian of a quantum Heisenberg
antiferromagnet  on a general lattice is  (in natural units)
\eq
H_{\mbox{\tiny AF}} = \sum_{\la i, j\ra} {\bf s_i} \cdot {\bf s_j} ,
\en
where the sum is over bonds $\la i, j \ra$ that connect sites $i$ and $j$
and ${\bf s} = (s^1, s^2, s^3)$ are spin operators in the spin-$s$
representation, where $s$ can be anything.
For the checkerboard lattice the hamiltonian is
up to a constant equal to half the sum 
of the total spin squared of all boxes (labelled by $x$)
\eq
H = \frac{1}{2}\sum_x ({\bf s_1} +{\bf s_2} + {\bf s_3}+ {\bf s_4})_x^2.
\label{checker}
\en

A 3 $\times$ 3 checkerboard with periodic boundary
conditions, i.e., with four independent sites,
provides the simplest example. It has
a hamiltonian that is (up to a constant) the total spin
squared of one box and the energy levels, degeneracies, and
eigenstates follow from the decompostion of $[s]^{\otimes 4}$ into
components of total spin; all ground states have total
spin zero and there are $2s +1$ of them.

A checkerboard lattice of arbitrary size,
with or without
periodic boundary conditions but with an even number of independent
sites, has the property
that it can be split into two equal parts
that are mirror images of one another about a line
that cuts bonds, as indicated in figure~\ref{fig:che1}, and that
contains no sites. We shall now show that such a system has at least
one spin-zero ground state.
It is actually not important, for the following argument,
what the lattice looks
like on the left or right; these sublattices neither need to
be checkerboards nor do they have to be purely antiferromagnetic 
(as long as total spin is a good quantum number).
What is impotant is, that the whole system is reflection symmetric about the
line that separates left and right and that
the crossing bonds are of checkerboard type. 
(For a system with periodic
boundary conditions in one direction there will 
actually be two such lines, but we
emphasize that PBC is not needed here even though it is needed in the usual
reflection positivity applications; see \cite{DLS} and references therein.)
A key observation is that these crossing bonds (solid lines in
figure~\ref{fig:che1}(b)) form antiferromagnetic bonds
$\bf S_L \cdot S_R$ between pairs of spins $\bf S_L = s_1 + s_2$ and 
$\bf S_R = s_3 + s_4$ of each box on the symmetry line.

The hamiltonian is $H = H_L + H_R + H_C$, where $H_L$ and $H_R$
act solely in the Hilbert spaces of the left, respectively right, subsystem
and $H_C$
contains the crossing bonds.
For the checkerboard
$H_C = \sum_y (({\bf s_1 + s_2})\cdot ({\bf s_3 + s_4}))_y$,
with the sum over boxes $y$ that are bisected by the
symmetry line.
$H_L$ and $H_R$ are completely arbitrary as long as they commute with
the total spin operator. 
We will, however, assume here that they are real in the $S^3$ basis.
Any state of the system can be written in terms of a 
matrix $c$ as
\eq
\psi = \sum_{\al, \be} c_{\al\be} 
\psi^L_\al \ot (\psi^R_\be)_{\mbox{\tiny rot}} , \label{psi}
\en
where the $\psi^L_\al$ form a real orthonormal basis of $S^3$ eigenstates
for the left subsystem and the $(\psi^R_\be)_{\mbox{\tiny rot}}$ are
the corresponding states for the right subsystem, 
but rotated by an angle $\pi$
around the 2-direction in spin space. This rotation takes $\uparrow$ into
$\downarrow$, $\downarrow$ into $-\uparrow$, and more generally
$|s,m\ra$ into $(-)^{s-m}|s,-m\ra$. It reverses the signs of the
operators $S^1$ and $S^3$, while it keeps $S^2$ unchanged.
The eigenvalue problem $H \psi = E \psi$ is now a matrix
equation
\eq
h_L c + c (h_R)^T -\sum_{i=1}^3 \sum_y t_y^{(i)} c (t_y^{(i)})^T = E c,
\label{ev}
\en
where $(h_L)_{\al\be}$ and $(h_R)_{\al\be}$ are real, symmetric matrix
elements of the corrosponding terms
in the hamiltonian and the $t_y^{(i)}$
are the \emph{real} matrices defined for the
spin operators $\bf s_1$ and $\bf s_2$
in box $y$ by
$t^{(1,3)}_{\al\be} = \la\psi_\al^L|s^{(1,3)}_1 + s^{(1,3)}_2|\psi_\be^L\ra$
and $t^{(2)}_{\al\be} = i \la\psi_\al^L|s^{(2)}_1 + s^{(2)}_2|\psi_\be^L\ra$.
Note the overall minus sign of the crossing term in (\ref{ev}):
replacing $\bf s_1 + s_2$ by $\bf s_3 + s_4$ and $\psi^L$ by
$(\psi^R)_{\mbox{\tiny rot}}$ gives a change of sign for directions 1 and 3,
while the $i$ in the definition of $t^{(2)}$ gives the minus sign 
for direction 2.

Consider now the energy expectation in terms of $c$:
\eq
\la\psi| H |\psi\ra =\tr c c^\dagger h_L + \tr c^\dagger c h_R
-\sum_{i,y} \tr c^\dagger t_y^{(i)} c (t_y^{(i)})^\dagger .
\label{ee}
\en
Since $H$ is left-right symmetric and by assumption real, we find that for an
eigenstate of $H$ with coefficient matrix $c$,
there is also an eigenstate with matrix $c^\dagger$ and, by linearity,
with $c+ c^\dagger$ and $i(c - c^\dagger)$. Without loss of generality we
may, therefore, write eigenstates of
$H$ in terms of Hermitean
$c = c^\dagger$.
We shall also take $\psi$ to be normalized:
$\la\psi|\psi\ra = \tr c^\dagger c = 1$.
Following \cite{L}, 
let us write the trace in the last term of (\ref{ee}) in the 
diagonal basis of $c$:
$-\tr c^\dagger t_y^{(i)} c (t_y^{(i)})^\dagger =
-\sum_{k,l} c_k c_l |(t_y^{(i)})_{kl}|^2$.
This expression clearly does not increase if we replace all the $c_k$
by their absolute values $|c_k|$, i.e., if we replace the matrix
$c$ by the positive semidefinite matrix $|c| = \sqrt{c^2}$.
The first two terms in (\ref{ee}) and the norm of $\psi$
remain unchanged under this operation.
We conclude that if $c$ is a ground state than so is $|c|$. Since
$c = c^+ - c^-$ and $|c| = c^+ + c^-$, with p.s.d.\ $c^+$ and $c^-$,
we may, in fact, chose a basis of ground states with p.s.d.\ coefficient
matrices.

Next, we will show that the state $\psi_0$ with the unit matrix as
coefficient matrix (in the $S^3$ eigenbasis) has total spin zero.
Since the overlap of $\psi_0$ with a state with matrix $c$ is simply the
trace of $c$, which is neccessarily non-zero for states with a p.s.d.\
matrix, and because spin is a good quantum number of the problem, this will
imply that there is a least one ground state with total spin zero.
First consider a spin-1/2 system. In the $S^3$ eigenbasis
every site has then either spin up or down. The state with unit coefficient
matrix is a tensor product of singlets on corresponding 
pairs of sites $i \in L$,
$i' \in R$ of the two sublattices:
\eq
\psi_0 = \bigotimes_{i \in L} \left(\uparrow (\uparrow)_{\mbox{\tiny rot}} + 
\downarrow (\downarrow)_{\mbox{\tiny rot}}\right)_{i i'} =
\bigotimes_{i \in L} \left(\uparrow \downarrow 
- \downarrow \uparrow\right)_{i i'} .
\en
The analogous state for a system with arbitrary spins,
\eq
\psi_0 
=\bigotimes_{i\in L} \sum_{m=-s}^s 
\left((-)^{s-m}|s,m\ra\otimes|s,-m\ra\right)_{i i'} ,
\en 
is also a tensor product of spin-zero states.

Finally, we would like to show that
the projection onto the spin zero part of a state with p.s.d.\
coefficient matrix preserves its positivity. This is only of academic
interest here, but it is non-trivial and may very well be important for
other physical questions.
To find the projection onto spin zero we
need to decompose the whole Hilbert space into tensor products of the
spin components
$[j]_k \ot [j']_{k'}$ of the two subsystems;
here $k$, $k'$ are additional quantum numbers that
distinguish multiple multiplets with the same spin $j$. Only tensor
products with $j = j'$ can have a spin zero subspace, which is
unique, in fact, and generated by the spin zero state
$\displaystyle\sum_{m=-j}^j |j,m,k\ra\ot (-)^{j-m} |j,-m,k'\ra$.
Noting that $(-)^{j-m}|j,-m,k'\ra$ is precisely the spin-rotated state
$(|j,m,k'\ra)_{\mbox{\tiny rot}}$, we convince
ourselves that the projection onto spin
zero amounts to a partial trace over $m$ in a suitably parametrized
matrix $c$.
This operation preserves positive semidefiniteness, so we actually proved that
the checkerboard has at least one ground state that has
both total spin zero \emph{and} a p.s.d. coefficient matrix $c$.

We do not know how many ground states there are.
To determine the spin of any remaining ground states
we add an external field to the hamiltonian and study the resulting
magnetization. We will see that the spontaneous magnetization of every box
on the symmetry line vanishes for all ground states, 
and thus if we have periodic
boundary conditions in at least one direction, 
the total magnetization vanishes. 
Since $S^3_{\mbox{\tiny tot}}$ is a good quantum number and
$S^\pm_{\mbox{\tiny tot}}$ commute with
the hamiltonian, this will imply that all ground states in such a system
have total spin zero. 
Let us thus modify the original hamiltonian (\ref{checker}) by  
replacing the term
$(s_1^{(3)} +s_2^{(3)} +s_3^{(3)} +s_4^{(3)})^2_z$
for a single box, $z$, on the symmetry line
by $(s_1^{(3)} +s_2^{(3)} +s_3^{(3)} +s_4^{(3)} - b)^2_z$, i.e.,
effectively adding a field $b$ to the spins in box $z$ 
and a constant term $\frac{1}{2}b^2$
to the hamiltonian. We want to distribute the resulting
$b$-terms $(s_1^{(3)} +s_2^{(3)} - b/2)^2$, $(s_3^{(3)} +s_4^{(3)} - b/2)^2$,
and $2 (s_1^{(3)} +s_2^{(3)} - b/2)(s_3^{(3)} +s_4^{(3)} - b/2)$
to $H_L$, $H_R$, $H_C$ respectively.
We cannot use the spin rotation as before, because
the crossing terms in the hamiltonian would no longer
be left-right symmetric in the basis (\ref{psi}). To avoid
this problem we will, instead, expand eigenstates $\psi$ 
in the same basis on the left and on the right:
\eq
\psi = \sum_{\al, \be} \tilde c_{\al\be} 
\psi^L_\al \ot \psi^R_\be.
\en
In this basis the hamiltonian is left-right symmetric and
we may assume, as before, that $\tilde c = \tilde c^\dagger$.
Except for the presence of $b$ in box $z$
the energy expectations on the left and right
are as before. The energy expectation of the crossing terms of box $z$
in the diagonal
basis of $\tilde c$ is now
$$
\sum_{k,l} \tilde c_k \tilde c_l \left(|(t_z^{(1)})_{kl}|^2
- |(t_z^{(2)})_{kl}|^2
+ |(t_z^{(3)})_{kl} - b/2|^2\right) .
$$
This expression clearly does not increase if we replace
the $c_k$ by their absolute value $|c_k|$ \emph{and}
change the signs of the first and last terms. The sign change
can be achieved by simultaneously performing a spin rotation
and changing the sign of the field $b$ in the right subsystem.
This actually completly removes $b$ from
the hamiltonian. We have thus shown that the ground state energies
of the systems $H_b$ with and $H_0$ without the $b$-terms satisfy the
inequality $E_b \geq E_0$. Let $|b\ra$ be a ground state of $H_b$
and $|0\ra$ a ground state of $H_0$. It follows from the variational
principle, that $\la 0|H_b|0\ra \geq \la b|H_b|b\ra = E_b \geq E_0$.
Expressed in terms of spin operators this reads
$E_0 - 2 \la 0|b (s_1^{(3)}+s_2^{(3)}+s_3^{(3)}+s_4^{(3)})_z|0 \ra
+ b^2 \geq E_0$.
Recalling that we are free to choose both the sign and
the magnitude of $b$ we find that the ground state magnetization of
box $z$ must be zero:
\eq
\la 0|(s_1^{(3)}+s_2^{(3)}+s_3^{(3)}+s_4^{(3)})_z|0\ra = 0.
\en
This quantum analog of the ``\emph{ice rule}''
is true for any box on the symmetry line and it holds 
for all three spin components. 
In a system with periodic
boundary conditions and an even number of sites in at least one direction
we can choose the symmetry line(s) to intersect any given box, so in such
a system the magnetization is zero both for every single 
box separately and also
for the whole system: $\la 0|S^{(3)}_{\mbox{\tiny tot}}|0\ra = 0$.
As mentioned previously this
implies that the total spin is zero for 
all ground states of such a system.

Let us return to the inequality $E_b \geq E_0$. It
implies a bound on the local susceptibility of the system: Let
$E(b) \equiv \la b|H_0 - b S^{(3)}_{\mbox{\tiny box}}|b\ra$ be the 
ground state energy of the periodic
pyrochlore checkerboard with a single box immersed in an external field
$b$. Recalling 
$H_b = H_0 - b S^{(3)}_{\mbox{\tiny box}} + \frac{1}{2} b^2$, we see
that the above inequality becomes $E(b) + \frac{1}{2} b^2 \geq E(0)$ and,
assuming differentiability, implies an
upper bound on the susceptibility at zero field for single-box magnetization
\eq
\chi_{\mbox{\tiny loc}} = 
-\frac{1}{\lambda} \left.\frac{\partial^2 E(b)}{\partial b^2}\right|_{b=0} 
\leq \frac{1}{4},
\en
where $\lambda = 4$ is the number of spins in a box. 
(The susceptibility is given
in natural units in which we have absorbed the g-factor 
and Bohr magneton in the definition 
of the field $b$.)

We would like to get more detailed information about the response of the
spin system to a global field $\{b_x\}$ in a hamiltonian
$H_{\{b_x\}}$ which is identical to (\ref{checker}), 
except for the terms for the third
spin component, which are replaced by 
$(s_1^{(3)}+s_2^{(3)}+s_3^{(3)}+s_4^{(3)} - b_x)_x^2$.
From what we have seen so far, it is apparent that the corresponding
ground state energy $E_{\{b_x\}}$
is extremal for $b_x =0$. With the help of a more sophisticated trace 
inequality \cite{KLS}, that becomes relevant 
whenever the matrix $c$ in (\ref{psi})
cannot be diagonalized, one can actually show that 
$E_{\{b_x\}}$ has an absolute minimum at
$b_x =0$:
\eq
E_{\{b_x\}} \geq E_0 . \label{ineq}
\en
Note that we had to put the field on the boxes 
for this result to hold; not every field 
on the individual spins can be written this way.
The special choice $b_x =B/2$ corresponds to a global homogenous field $B$ on
all spins. (The factor $1/2$ adjusts for the fact that every spin is shared by
two boxes.) If 
$E(B) = \la B|H_0 - B S^{(3)}_{\mbox{\tiny tot}}|B\ra$ is the ground
state energy of the periodic pyrochlore checkerboard 
in the external field $B$, 
then (\ref{ineq}) implies $E(B) + \frac{\Lambda}{16} B^2 \geq E(0)$, 
and thus an upper
bound on the susceptibility per site at zero field (in natural units)
\eq
\chi = 
-\frac{1}{\Lambda}\left.\frac{\partial^2 E(B)}{\partial B^2}\right|_{B=0} 
\leq \frac{1}{8} ,
\en
where $\Lambda$ is the number of independent sites,
which equals twice the number of boxes.

All these results continue to hold at finite temperature. 
The analog of (\ref{ineq})
holds also for the partition function corresponding to $H_{\{b_x\}}$:
\eq
Z_{\{b_x\}} \leq Z_0 , \label{ineqz}
\en
as can be shown by a straightforward application of lemma~4.1 
in section 4 of \cite{DLS}
to the pyrochlore checkerboard. The physically relevant partition function
for the periodic pyrochlore checkerboard at finite temperature in a homogenous
external field, 
$Z(B) \equiv \tr e^{-\beta(H_0 - B S^{(3)}_{\mbox{\tiny tot}})}$,
differs from $Z_{\{b_x\}}$, where $b_x = B/2$, 
only by a factor corresponding to the 
constant term in $H_{\{b_x\}}$. Due to (\ref{ineqz}), the free energy 
$F(B)= -\beta^{-1}\ln Z(B)$ satisfies
\eq
F(B) + \frac{\Lambda}{16} B^2 \geq F(0).
\en
This implies (i) that the magnetization at zero field is still zero at finite
temperature,
\eq
M_T = 
-\frac{1}{\Lambda}\left.\frac{\partial F(B)}{\partial B}\right|_{B=0} = 0,
\en
and, more interestingly, (ii) the same upper bound for the susceptibility 
per site at zero field as we had for the ground state:
\eq
\chi_T = 
-\frac{1}{\Lambda}\left.\frac{\partial^2 F(B)}{\partial B^2}\right|_{B=0} 
\leq \frac{1}{8} .
\en
The bounds on the susceptibility hold for arbitrary intrinsic spin-$s$
and agree very well with the results of \cite{MB} for the classical
pyrochlore antiferromagnet in the un-diluted case.

It is not essential for our method that only every other square of the
pyrochlore checkerboard is a frustrated unit, 
only the reflection symmmetry and the
antiferromagnetic crossing bonds are important. We could, e.g., have
diagonal bonds on \emph{every} square, but then the horizontal and/or
vertical bonds must have twice the coupling strength.
Our results also apply to  various 3-dimensional cubic
versions of the checkerboard, e.g., 
with diagonal crossing bonds in every other
cube. While the method does not directly work for the 3D pyrochlore lattice
because its geometry is too complicated, it has been seen in \cite{MC}
that classically this system has similar properties to the pyrochlore
checkerboard, which is also fully frustrated, and has the added advantage
of being more easily visualizeable. 

We would like to thank Roderich Moessner 
for bringing this problem to our attention,
for explaining his work on the classical pyrochlore antiferromagnet,
and for numerous helpful discussions and suggestions.


\begin{references}

\bibitem{HTD}
For a collection of articles, see 
{\em Magnetic systems with competing interactions: 
frustrated spin systems}, 
edited by H. T. Diep (World Scientific, Singapore, 1994); 
for Ising systems, see:                  
R. Liebmann, {\em Statistical Mechanics of Periodic Frustrated Ising
Systems} (Springer, Berlin, 1986).

\bibitem{WH} G. H. Wannier, \ti{Phys. Rev.}{79}{357}{1950}{},
erratum: \ti{\prb}{7}{5017}{1973}{};
R. M. F. Houtappel, \tit{Physica}{16}{425}{1950}{}

\bibitem{PWA}
P. W. Anderson, \tit{Phys. Rev.}{102}{1008}{1956}{Ordering and
Antiferromagnetism in Ferrites}

\bibitem{JV}
J. Villain, \tit{\zpb}{33}{31}{1979}{Insulating Spin Glasses}

\bibitem{MC} 
R. Moessner and J. T. Chalker,
\ti{\prl}{80}{2929}{1998}{Properties of a classical spin liquid: The
Heisenberg pyrochlore antiferromagnet};
\tit{\prb}{58}{12049}{1998}{Low-temperature properties of classical
geometrically frustrated antiferromagnets}


\bibitem{SL}%
R. R. Sobral and C. Lacroix, \tit{\ssc}{103}{407}{1997}{Order by
disorder in the pyrochlore antiferromagnets} 


\bibitem{CL}
B. Canals and C. Lacroix, \tit{\prl}{80}{2933}{1998}{Pyrochlore
antiferromagnet: A three-dimensional quantum spin liquid} 

\bibitem{RVB} M. Isoda, S. Mori, 
\tit{J. Phys. Soc. Jap.}{67}{4022}{1998}{Valence-bond crystal and anisotropic
excitation spectrum on 3-dimensional frustrated pyrochlore}

\bibitem{LM} E. H. Lieb, D. Mattis, J. Math. Phys. {\bf 3}, 749 (1962).

\bibitem{DLS} F. J. Dyson, E. H. Lieb, B. Simon, J. Stat. Phys. {\bf 18},
335 (1978).




\bibitem{L} E. H. Lieb, Phys. Rev. Lett. {\bf 62}, 1201 (1989).

\bibitem{MB} R. Moessner, A. J. Berlinsky, 
\mbox{cond-mat/}9906421 (1999).

\bibitem{KLS} T. Kennedy, E. H. Lieb, B. S. Shastry, J. Stat. Phys. 
{\bf 53}, 1019 (1988); E. H. Lieb, P. Schupp, unpublished.



\end{references}
\end{document}